\documentclass[conference]{IEEEtran}
\IEEEoverridecommandlockouts
\usepackage{cite}
\usepackage{amsmath,amssymb,amsfonts}
\usepackage{algorithmic}
\usepackage{graphicx}
\usepackage{textcomp}
\usepackage{xcolor}
\usepackage{placeins}
\usepackage{tabularx}

\def\BibTeX{{\rm B\kern-.05em{\sc i\kern-.025em b}\kern-.08em
    T\kern-.1667em\lower.7ex\hbox{E}\kern-.125emX}}
\begin{document}

\title{A Resource-Driven Approach for Implementing CNNs on FPGAs Using Adaptive IPs\\

}

\author{\IEEEauthorblockN{1\textsuperscript{rd} Philippe MAGALHÃES}
\IEEEauthorblockA{\textit{Lab. H. Curien, UMR 5516} \\
\textit{CNRS, IOGS, Univ. J. Monnet}\\
Saint Etienne, France\\
philippe.magalhaes@univ-st-etienne.fr}
\and
\IEEEauthorblockN{2\textsuperscript{rd} Virginie FRESSE}
\IEEEauthorblockA{\textit{Lab. H. Curien, UMR 5516} \\
\textit{CNRS, IOGS, Univ. J. Monnet}\\
Saint Etienne, France\\
virginie.fresse@univ-st-etienne.fr}
\and
\IEEEauthorblockN{3\textsuperscript{rd} Benoît SUFFRAN}
\IEEEauthorblockA{\textit{ST Microelectronics} \\
Grenoble, France\\
benoit.suffran@st.com}
\and

\IEEEauthorblockN{4\textsuperscript{rd} Olivier ALATA}
\IEEEauthorblockA{\textit{Lab. H. Curien, UMR 5516} \\
\textit{CNRS, IOGS, Univ. J. Monnet}\\
Saint Etienne, France\\
olivier.alata@univ-st-etienne.fr}
\and

}

\maketitle

\begin{abstract}
The increasing demand for real-time, low-latency artificial intelligence applications has propelled the use of Field-Programmable Gate Arrays (FPGAs) for Convolutional Neural Network (CNN) implementations. FPGAs offer reconfigurability, energy efficiency, and performance advantages over GPUs, making them suitable for edge devices and embedded systems. This work presents a novel library of resource-efficient convolution IPs designed to automatically adapt to the available FPGA resources. Developed in VHDL, these IPs are parameterizable and utilize fixed-point arithmetic for optimal performance. Four IPs are introduced, each tailored to specific resource constraints, offering flexibility in DSP usage, logic consumption, and precision. Experimental results on a Zynq UltraScale+ FPGA highlight the trade-offs between performance and resource usage. The comparison with recent FPGA-based CNN acceleration techniques emphasizes the versatility and independence of this approach from specific FPGA architectures or technological advancements. Future work will expand the library to include pooling and activation functions, enabling broader applicability and integration into CNN frameworks.

\end{abstract}

\begin{IEEEkeywords}
FPGA, CNN, Optimization, Adaptation
\end{IEEEkeywords}

\section{Introduction}

The evolution of artificial intelligence (AI) and neural networks integrates advances in statistical algorithms and brain-inspired models. Breakthroughs such as backpropagation in the 1980s revolutionized multilayer networks, while Convolutional Neural Networks (CNNs) gained prominence with LeNet and AlexNet \cite{shi}. Today, CNNs are essential in machine learning tasks like image recognition and object detection~\cite{seng}.

To meet the growing demand for efficiency and performance in edge devices and embedded systems, FPGAs have emerged as a viable alternative to GPUs. While GPUs excel in high-performance parallel processing for neural networks  and consolidated support, their high power consumption and latency pose challenges. In contrast, FPGAs offer reconfigurability, energy efficiency, and low latency, making them ideal for real-time and highly customized applications~\cite{liu}

FPGAs combine a wide range of resources for logical, arithmetic, and storage operations, making them ideal for customized and high-performance applications. Configurable Logic Blocks (CLBs) form the core of FPGAs, containing Look-Up Tables (LUTs), flip-flops, and multiplexers to implement logical and sequential functions. These CLBs are organized into slices, with SliceL optimized for combinational logic and SliceM offering additional support for distributed memory and shift registers. Arithmetic operations are accelerated by carry chains, which efficiently propagate carry signals, essential for fast calculations in adders, multipliers, and accumulators. FPGAs also include memory blocks, such as BRAM for temporary data storage, UltraRAM for higher capacity in modern devices, and distributed memory configured within LUTs for local data. Additionally, Digital Signal Processing (DSP) units handle intensive operations like multiplications and accumulations, which are critical for CNN applications. These resources are interconnected by programmable routing networks and switch matrices, ensuring flexibility in communication between blocks. Optimizing the use of these resources is crucial to maximize the performance and energy efficiency of CNN implementations. strategies to improve the use of these resources are fundamental in this context \cite{seng}.

Recent studies have proposed a wide range of techniques to optimize CNNs on FPGAs. Work such as \cite{luo} highlights the use of pipelined architectures and the exploitation of parallelism. \cite{shao} demonstrates that quantizing network weights and data to smaller bit-widths can significantly reduce the use of DSPs and BRAMs. The systolic array architectures implemented by \cite{basalema} use local communication and regular layout, which allows achieving high clock frequency and reducing global data communication. \cite{shi} uses dynamic partial reconfiguration to increased logic capacity, keeping resources in use and freeing up space from idle resources. These and other techniques can maximize throughput and minimize latency. However, they often require significant resources, reduce accuracy, are restricted to a specific architecture, or cause hardware overhead.

This work distinguishes itself by prioritizing the efficient utilization of FPGA resources, independent of the architecture or technological advancements. A library of Intellectual Properties (IPs) has been developed to enable automatic adaptation to the available resources, ensure hardware independence, provide scalability, and emphasize balanced resource allocation.

\section{Proposed IP library for convolution}

We designed four convolution IPs in VHDL, each tailored to specific resource constraints and computational needs. All IPs are parameterizable, using fixed-point arithmetic for efficiency. The kernel coefficients are loaded serially to optimize memory usage, while data inputs are loaded in parallel for improved throughput. Table~\ref{tab:convolution_ips} summarizes the characteristics of each convolution IP.

\begin{table}[h]
\centering
\caption{Characteristics of Developed Convolution IPs}
\label{tab:convolution_ips}
\begin{tabular}{|p{0.8 cm}|p{1.4 cm}|p{1.5 cm}|p{3 cm}|}
\hline
\textbf{IP}     & \textbf{DSP Usage}      & \textbf{Logic Usage}          & \textbf{Key Features}                               \\ \hline
$Conv_1$         & None                    & High                          & Only logic, no DSP;  \\ 
                 &                         &                               & one convolution per cycle. \\ \hline
$Conv_2$         & 1 DSP                   & Moderate                      & Reduces the use of logic; \\ 
                 &                         &                               & one convolution per cycle. \\ \hline
$Conv_3$         & 1 DSP                   & High                          & Two parallel convolutions; \\ 
                 &                         &                               & limited up to 8-bit operands. \\ \hline
$Conv_4$         & 2 DSPs                  & Moderate                      & Two parallel convolutions; \\ 
                 &                         &                               & optimized for parallelism. \\ \hline
\end{tabular}
\end{table}

\section{Experiments and Results}
\subsection{Resource Utilization}

Experiments were conducted using Xilinx Vivado on a Zynq UltraScale+ ZCU104 at 200 MHz, with 8-bit fixed-point data and a 3x3 kernel. Table~\ref{tab:resource_utilization} summarizes the utilization of resources of each convolution IP, highlighting trade-offs between DSP usage, logic consumption, and performance.

\begin{table}[h]
\centering
\caption{Resource Utilization of Convolution IPs}
\label{tab:resource_utilization}
\begin{tabular}{|p{0.8 cm}|p{0.55 cm}|p{0.55 cm}|p{0.55 cm}|p{0.55 cm}|p{1.2 cm}|p{1.3cm}|}
\hline
\textbf{IP} & \textbf{LUTs} & \textbf{Regs} & \textbf{CLBs} & \textbf{DSPs} & \textbf{WNS (ns)} & \textbf{Power (W)} \\ \hline
$Conv_1$     & 105           & 54            & 15            & 0             & 2.596             & 0.593             \\ \hline
$Conv_2$     & 30            & 22            & 5             & 1             & 2.276             & 0.594             \\ \hline
$Conv_3$     & 45            & 32            & 10            & 1             & 2.086             & 0.594             \\ \hline
$Conv_4$     & 42            & 23            & 8             & 2             & 2.870             & 0.596             \\ \hline
\end{tabular}
\end{table}

Conv\_1 consumes high logical resources, but is suitable for FPGAs with limited DSPs. Conv\_2 uses one DSP, significantly reducing logic usage. It is ideal for FPGAs with DSP availability and limited logic resources. Conv\_3 Performs two simultaneous convolutions using a single DSP, suitable for applications that require greater parallelism with minimal use of DSPs. Limits operands to 8 bits, resulting in reduced precision. Conv\_4 Uses two DSPs to execute two convolutions in parallel, intended for scenarios that demand high parallelism and have wide availability of DSPs. Provides greater precision by allowing larger operands. The experimental results highlight the differences between the IPs and reinforce the purpose of each of them.

\subsection{Timing and Routing Congestion}

All IPs meet timing constraints with positive Worst Negative Slack (WNS) values. Conv\_4 exhibits the highest timing robustness, while Conv\_3 demonstrates the lowest due to its increased complexity. No routing congestion issues were observed during the analysis.

\section{Comparison with Related Works}

Table~\ref{tab:comparison} compares this work with other recent approaches, highlighting differences in focus and resource adaptability. This comparison highlights the unique strengths of this work, particularly its ability to balance resource efficiency, scalability, and hardware independence. Unlike other approaches, the proposed IP library adapts seamlessly to diverse resource constraints, offering a robust and versatile solution for CNN deployment on FPGAs.

\begin{table}[!ht]
\centering
\caption{Comparison of Optimization Techniques for CNNs on FPGAs}
\label{tab:comparison}
\begin{tabularx}{\columnwidth}{|p{40pt}|p{40pt}|p{35pt}|p{35pt}|X|}
\hline
\textbf{Attribute}         & \textbf{This Work}         & \textbf{Luo et al.~\cite{luo}} & \textbf{Shao et al.~\cite{shao}} & \textbf{Shi et al.~\cite{shi}} \\ \hline
Focus                           & Adaptation                   & Maximize                       & Maximize                        & Optimize                   \\
                                & to resources                 & throughput                     & throughput                      & Resource                   \\ \hline
FPGA Architecture Dependency    & Low                        & High                           & High                            & Medium                     \\ \hline

Multiple Precisions             & Yes                        & Yes                            & Yes                             & No                         \\ \hline

Model                           & High                       & Medium                         & Medium                          & High                       \\ 
Scalability                     &                            &                                &                                 &                            \\ \hline
Resource Flexibility            & High                       & Low                            & Low                             & Medium                     \\ \hline
\end{tabularx}
\end{table}

\section{Conclusion}

This work introduces a novel library of convolution IPs designed for FPGA-based CNN deployment, prioritizing scalability, flexibility of resource usage, and independence from specific hardware advancements. The experimental results demonstrate the effectiveness of the proposed IPs in balancing logic and DSP usage, making them suitable for a wide range of applications. Compared to recent approaches, this work stands out by offering a robust and adaptable solution that accommodates diverse FPGA configurations, emphasizing resource efficiency without compromising scalability. Future efforts will focus on expanding the IP library to support additional CNN layers, automating IP selection based on resource availability, and integrating these designs for real-world applications.

\section *{Acknowledgments}

This work was sponsored by a public grant overseen by Auvergne-Rhône-Alpes region, Grenoble Alpes Metropole and BPIFrance, as part of project I-Démo Région "Green AI".

\end{document}